\documentclass[preprint,superscriptaddress,floatfix,sort&compress]{revtex4}

\usepackage[dvips]{graphicx}
\usepackage{amssymb,amsfonts,amsmath}
\usepackage{color}
\usepackage[font=scriptsize]{caption}

\usepackage{ulem}

\begin{document}

\title{Non-Equilibrium Effects of Molecular Motors on Polymers}

\author{M. Foglino}
\affiliation{SUPA, School of Physics and Astronomy, University of Edinburgh, Edinburgh EH9 3FD, UK}
\author{E. Locatelli}
\affiliation{Faculty of Physics, University of Vienna, Boltzmanngasse 5, A-1090, Vienna, Austria}
\author{C. A. Brackley}
\affiliation{SUPA, School of Physics and Astronomy, University of Edinburgh, Edinburgh EH9 3FD, UK}
\author{D. Michieletto}
\affiliation{SUPA, School of Physics and Astronomy, University of Edinburgh, Edinburgh EH9 3FD, UK}
\author{C. Likos}
\affiliation{Faculty of Physics, University of Vienna, Boltzmanngasse 5, A-1090, Vienna, Austria}
\author{D. Marenduzzo}
\affiliation{SUPA, School of Physics and Astronomy, University of Edinburgh, Edinburgh EH9 3FD, UK}

\begin{abstract}
We present a generic coarse-grained model to describe molecular motors acting on polymer substrates, mimicking, for example,  RNA polymerase on DNA or kinesin on microtubules. The polymer is modeled as a connected chain of beads; motors are represented as freely diffusing beads which, upon encountering the substrate, bind to it through a short-ranged attractive potential. When bound, motors and polymer beads experience an equal and opposite active force, directed tangential to the polymer; this leads to motion of the motors along the polymer contour. The inclusion of explicit motors differentiates our model from other recent active polymer models. We study, by means of Langevin dynamics simulations, the effect of the motor activity on both the conformational and dynamical properties of the substrate. We find that activity leads, in addition to the expected enhancement of polymer diffusion, to an effective reduction of its persistence length. We discover that this effective ``softening'' is a consequence of the emergence of double-folded branches, or hairpins, and that it can be tuned by changing the number of motors or the force they generate. Finally, we investigate the effect of the motors on the probability of knot formation. Counter-intuitively our simulations reveal that, even though at equilibrium a more flexible substrate would show an increased knotting probability, motor activity leads to a marked \textit{decrease} in the occurrence of knotted conformations with respect to equilibrium.
\end{abstract}

\maketitle

	\section{Introduction}

Molecular motors that work out-of-equilibrium are key elements for the viability of cells. Typical examples include myosin and kinesin motors~\cite{Sanchez2012nature}, RNA and DNA polymerases~\cite{Alberts2014}, helicases~\cite{Calladine1997} and condensin~\cite{Le2013,Terakawa2017}. Although many of these protein complexes are biochemically well characterised, we are far from having a full understanding of their collective action \emph{in vivo}. In some cases, the behaviour may be tuned by inter-motor interactions that are mediated by the substrate itself, a germane example being the change in polymerase-DNA binding affinity due to substrate supercoiling~\cite{Brackley2016supercoil,Naughton2013}.   

Most previous theoretical and computational work on non-equilibrium forces within live cells has focused on modelling the propagation of stresses on biopolymer networks~\cite{Lenz2014,Broedersz2014} or active fields which mimic cytoskeleton dynamics~\cite{Tjhung2012}. 
While molecular motors can now be readily investigated using single-molecule techniques~\cite{Herbert2008,michaelis2013single,Ganji2018}, the emergent properties of systems where a collection of motors interact on, and through, their target polymer substrate remain poorly studied~\cite{bianco_loca,SaintillanMotor}.

Here we propose a coarse-grained computational model to describe the action of generic molecular motors on polymers with the aim of improving our understanding of the effects of ATP-consuming translocating machineries on the conformational and dynamic properties of their target substrates. Within our model, motors undergo free diffusion in 3-D {until they encounter the polymer, at which point they become ``bound'' to it via an attractive interaction.} {While bound, a motor experiences an active force propelling it along the polymer; at the same time the polymer is subjected to an equal but opposite active force. The result is that the motor displays relative motion with respect to the substrate. The motor can unbind from the polymer, for example due to thermal motion or because it reaches the polymer's end.}  
While our model is generic, and our aim is to understand its underlying physics, it could be viewed as a model for DNA or RNA polymerases moving along DNA or kinesin molecules stepping on microtubules~\cite{Alberts2014}.

{There has been a great deal of recent simulation work exploring active polymers, and our model differs from these in a number of ways. To our knowledge our work is the first to explicitly treat discrete motors that track along a polymer in a fully 3-D system. A number of recent studies~\cite{Harder2014,Jiang2014,Isele-Holder2015,Eisenstecken2016,bianco_loca,Anand2018,SaintillanMotor} have considered a chain of beads, where each bead experiences an active force, either directed along the chain bonds~\cite{Jiang2014,Isele-Holder2015,Anand2018} or in a direction independent of the chain~\cite{Eisenstecken2016,bianco_loca}. These works have mainly focused on regimes of different behaviour at different levels of activity (e.g. transitions from active translation to rotational motion~\cite{Jiang2014} or to spiral configurations~\cite{Isele-Holder2015} have been observed -- for a review see Ref.~\cite{Winkler2017}). In the present work the polymer only experiences an active force at the points where the explicit motors are bound (and these vary in time), and we focus on the effect on the conformational and dynamical properties of the polymer. Another class of models have more specifically focused on motor proteins found in the cytoskeleton, like myosin and kinesin which have provided a model experimental system for active materials~\cite{Schaller2010,Sanchez2012}. These can be modelled at a field theoretic level, but there have been a number of simulation studies which resolved the motors explicitly~\cite{nedelec2007,Freedman2017,Ravichandran2017,Gupta2019}. Unlike the current work, those studies focused on dense suspensions of motors and their substrates, and/or 2-D systems where the motors are fixed to a surface, and have often considered joined pairs of motors which can crosslink different substrate molecules to drive them into relative motion. In contrast, here we consider a single, semi-flexible polymer which is free to move in 3-D, and is subject to the action of multiple motors. }

Our simulations reveal that motor activity affects the steady state conformations of the polymer substrate; we find that these are more crumpled 
with respect to their passive counterparts. We attribute this effective softening of the substrate to the emergence of double-folded branches, or hairpins; the effect can be tuned by changing, e.g., the strength of motor activity. We also find that, for large enough motor activity, the centre of mass of the polymer substrate can display super-diffusive behaviour at short times, with a return to diffusive behaviour at long times. Again this effect can be controlled by varying the activity or number of motors. 
We conclude our work by investigating the effect of motors on 
the propensity of the polymer to form knots. 
This is relevant for many biological processes involving DNA, where knotting would be detrimental (e.g knots and tangles might hinder transcription and replication, repair of double strand breaks, or segregation of chromosomes).
We find that the motor activity effectively \textit{reduces} the steady state probability of finding knotted conformations. This is surprising since at equilibrium, in a good solvent, and in the absence of confinement, softer substrates show an enhanced probability of knotting~\cite{Coronel2017}. Thus, our findings point to an intriguing non-equilibrium effect of molecular motors that could be important in many biological systems. 

\begin{figure}[t!]
	\centering
	\includegraphics[width=.45\textwidth]{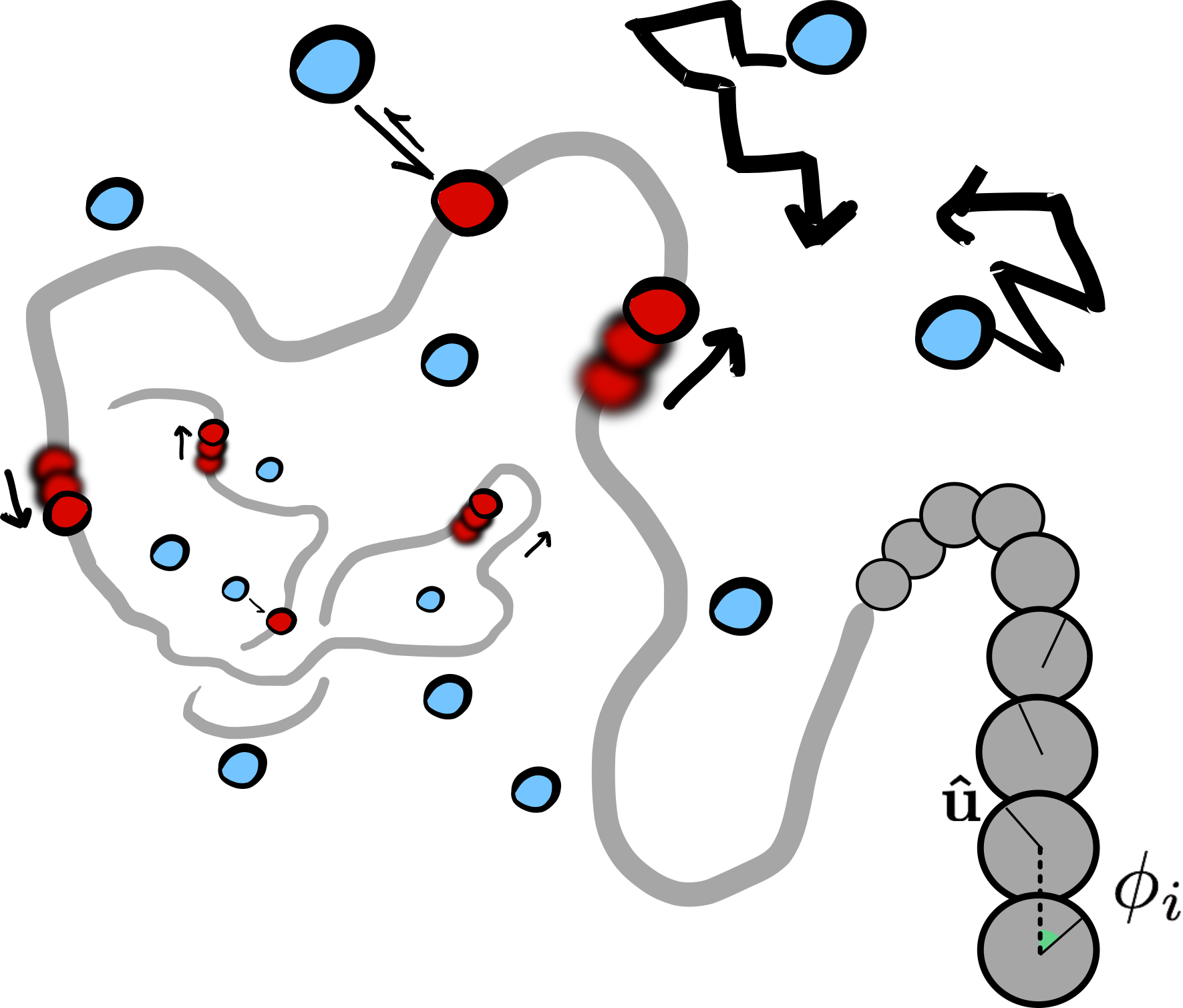}
	\caption{Schematic of our model for molecular motors acting on a generic polymer substrate. Diffusing motors (blue) can interact with the polymer backbone (grey); when bound, the motors (red) and the polymer experience opposite forces which cause relative motion. On the right hand side of the figure we show the polymer in more detail: the vector $\mathbf{\hat{u}}$ defines an orientation for each bead and $\phi_i$ is then defined as the angle between $\mathbf{\hat{u}}_i$ and the vector joining beads $i$ and $i+1$; an orientation potential acts to minimise $\phi_i$ (see text and Appendix~\ref{model_details} for details).}
	\label{fig:model}
\end{figure}

\section{The Model}
We perform Langevin dynamics simulations of a semi-flexible bead-and-spring polymer consisting of $L$ beads in solution with $N$ motors, as described schematically in Fig.~\ref{fig:model}. We use a standard polymer model which includes finitely-extensible non-linear elastic (FENE) springs connecting consecutive beads, a Kratky-Porod potential for triplets of beads providing bending rigidity, and a Weeks-Chandler-Anderson (WCA) potential to regulate steric interactions (see Appendix \ref{model_details} for full details). The intrinsic stiffness of the polymer is controlled by the bending energy $K_{\rm BEND}$ which appears as a parameter in the Kratky-Porod potential. Each polymer bead also displays an orientation, which is tracked by a vector $\mathbf{\hat{u}}_i$ as shown in Fig.~\ref{fig:model}, and {we include} a potential to orient this vector along the local direction of the polymer backbone {(if the position of bead $i$ is $\mathbf{r}_i$ and the orientation is $\hat{\mathbf{u}}_i$, then the potential acts to minimise the angle $\phi_i$ between $\hat{\mathbf{u}}_i$ and $|\mathbf{r}_{i+1}-\mathbf{r}_i|$)}.


The motors are represented by freely diffusing beads which interact with each other sterically (via a WCA potential) and via a Morse-like attractive {potential with the polymer beads. The latter is given by}
\begin{align}{
\label{motorattraction}
U_{\rm MOT}(r) = \left\{ 
\begin{array}{ll} 
K_{\rm MOT} \left[ (e^{-4r} - 2e^{-2r}) - \right.&  \\ 
\left. ~~~~ (e^{-4r_{\rm m\,cut}} - 2e^{-2r_{\rm m\,cut}}) \right], & r\leq r_{\rm m\,cut} \\
0, & \mbox{otherwise},
\end{array} \right.}
\end{align}
{Here $r$ is the the separation of the motor and the polymer bead, and $K_{\rm MOT}$ is the interaction strength.} {The potential} has a minimum when the motor and polymer beads overlap. {We consider a motor to be ``bound'' to a polymer bead when the centre-to-centre separation of the bead and the motor is less than $r_{\rm m\,cut}$, which we set to be the same as the motor diameter.}
In addition to this attraction, when a motor is bound to a polymer bead, it experiences an active force of magnitude $f$ {in the direction of the} orientation vector $\hat{\bold{u}}_i$ of the bead{; the polymer bead experiences an opposite active force $-f\hat{\mathbf{u}}_i$ (in compliance with Newton's third law).} {It is possible for a motor to be within a distance $r_{\rm m\, cut}$ of more than one polymer bead at the same time: in this case each interacting polymer bead will experience an active force, and the total active force on the motor will be the sum of that due to each interacting polymer bead. (In practice motors spend over 95\% of their ``bound'' time interacting with two consecutive polymer beads.)}

The result {of the active force is that the motors and the polymer are driven into relative motion; due to the connectivity of the chain, the polymer tends to move less compared to motors, so the latter effectively ``track'' along the polymer substrate. Motors can become unbound from the polymer due to thermal fluctuations (controlled by the parameter $K_{\rm MOT}$), and also active effects (e.g. tracking off the end of the polymer -- this is discussed in mote detail below).}
Since {each monomer has a preferred orientation, i.e.} the polymer is polar, motors {will} move in the parallel or anti-parallel direction depending on the sign of $f$. In the present work we consider only the case where all motors are the same and $f\geq0$.  

For simplicity motors and polymer beads are taken to have the same {diameter}, $\sigma$, mass, $m$, and unless otherwise stated they experience the same friction $\xi$ due to the implicit solvent {(though in reality a motor such as RNA polymerase would be much bigger)}. 
{We use the LAMMPs molecular dynamics software~\cite{LAMMPS} to perform the simulations: a Langevin equation for each polymer bead and motor is solved using a velocity-Verlet algorithm. A separate Langevin equation is solved for each monomer orientation $\hat{\mathbf{u}}_i$. Full details are given in Appendix \ref{model_details}. Throughout we quote energies and times in units of $k_BT$ and $\tau_{\rm LJ}=\sqrt{m\sigma^2/k_BT}$ respectively; how these map to physical units will depend on the system of interest, as detailed in Appendix \ref{model_details}.}

\section{Results}


{As discussed above, although a motor and the polymer bead it is bound to experience forces of equal magnitude (but acting in opposite directions), the chain connectivity means that the motors can move more freely, and tend to track along the polymer substrate.} 
{The three control parameters which we vary in our simulations are the number of motors $N$, the interaction strength $K_{\rm MOT}$, and the force acting on the motors $f$. However, in what follows we also consider $\langle n \rangle$, corresponding to the mean number of motors bound to the polymer at any one time. We note that these parameters are not independent: for large forces, motors have a shorter residency time due to the finite size of the polymer substrate and if both $N$ and $f$ are large, collisions between motors can also lead to a decrease in residency time. That is to say, changing $f$ at fixed $N$ leads to a change in $\langle n \rangle$, and this relationship depends on $N$ itself -- this is examined in more detail in Appendix \ref{ssec:motors}.} 

{As expected, the tracking speed of the motors in general increases with $f$, though the relationship is complex (as noted, changing $f$ leads to a change in $\langle n \rangle$ which can affect motor speed due to collisions). Within the range of parameters investigated, typically a motor will move from one polymer bead to the next in a few $\tau_{\rm LJ}$ (e.g. for $f=20$ and $N=200$ we measure a speed $\sim0.4$~beads/$\tau_{\rm LJ}$). For an $L=500$ bead polymer, motors typically spend on the order 50--100~$\tau_{\rm LJ}$ bound to it (increasing with $K_{\rm MOT}$ and decreasing with increasing $f$ and $\langle n \rangle$).}

{The motor action also changes the dynamics of the polymer. At a global scale this can be quantified by measuring the decorrelation time for the polymer end-to-end vector. For an $L=500$ bead polymer in the absence of motor action ($f=0$) a typical decorrelation time is of the order $10^5~\tau_{\rm LJ}$. Again, there is a complicated dependence on $f$ and $N$, but the motor effect leads to a \textit{decrease} in this decorrelation time by around an order of magnitude. Nevertheless, the polymer relaxation time remains much longer than the motor time-scales detailed above, i.e., the positions of motors on the polymer change quickly with respect to the polymer dynamics. This implies that, after some transient time, the properties of the polymer will settle into a non-equilibrium steady-state (NESS) which is independent of initial conditions (and the system is ergodic). In the following sections we examine these steady state polymer properties by averaging measurements over time and an ensemble of repeat simulations (50 independent simulations unless otherwise stated).}

\subsection*{Motor activity leads to polymer softening} 

\begin{figure*}[h!]
	\centering
	\includegraphics{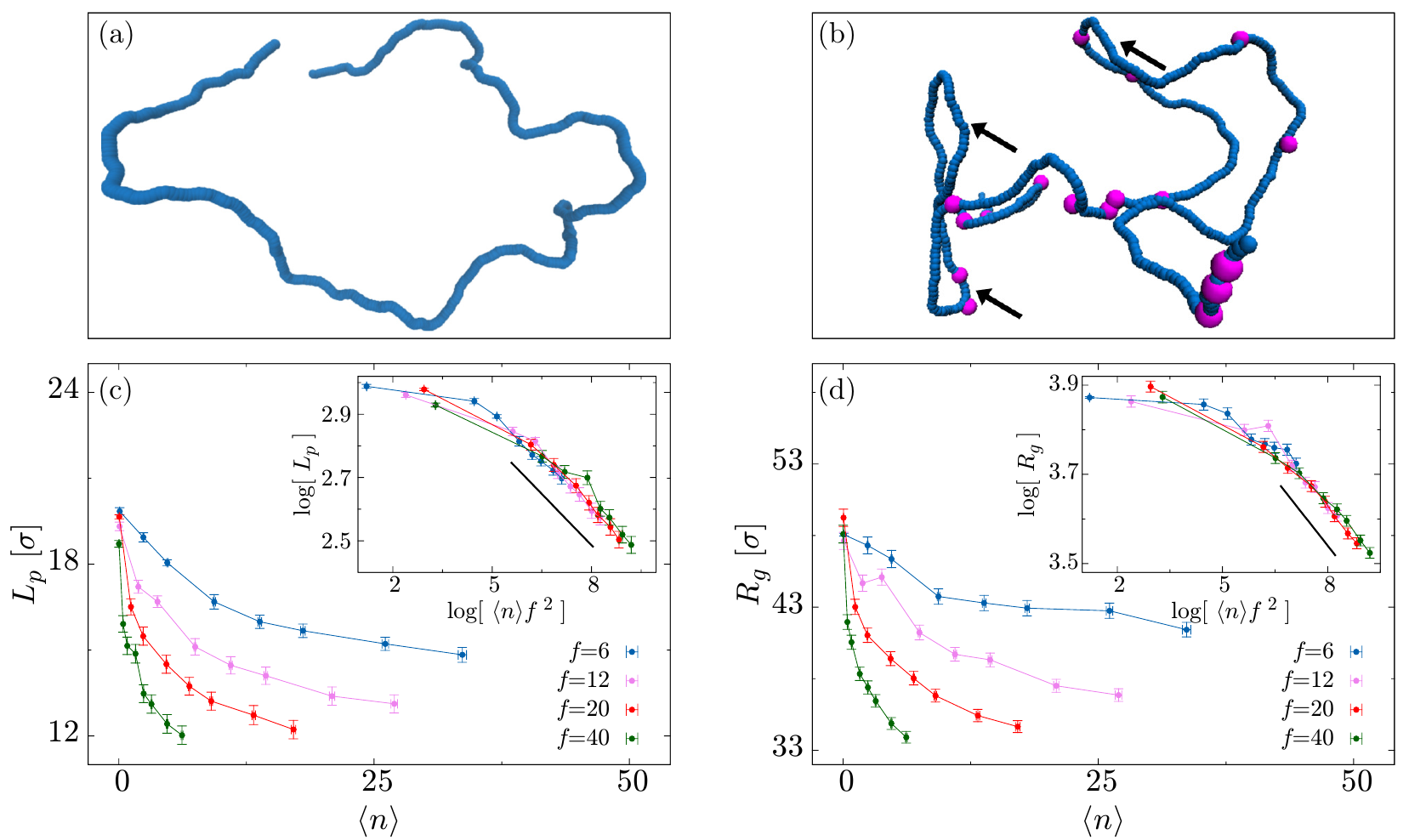}
	\caption{\small \textbf{(a-b)}  Comparison of conformations for a polymer of length $L =$ 500 with $\langle n \rangle=0$ and $\langle n \rangle=17$ motors attached at any time. In (b) hairpins are indicated by arrows and the purple beads mark the presence of an active motor bound to the polymer backbone. For clarity of visualisation, motors are depicted as having a bigger size with respect to the polymer beads. \textbf{(c)} Persistence length $L_p$ as a function of the average number of motors attached $\langle n \rangle$ for different values of $f$. Inset: the persistence length for different motor forces can be scaled on a master curve $L_p(x=\langle n \rangle f^2)\sim x^\alpha$ with $\alpha = -0.108 \pm  0.007$. \textbf{(d)} Similar plot showing $R_g$ as a function of $\langle n \rangle$. Again the inset shows the collapse of the curves when plotted as a function of $\langle n\rangle f^2$ with coefficient $\alpha = -0.100 \pm 0.009$. Error bars show the standard error in the mean (though these are often smaller than the points), and the lines connecting symbols are always a guide for the eye.}
	\label{fig:snaps}
\end{figure*}

{In this section we study} the role of motors on the steady state conformations of the polymer substrate.  We examine the effect of different values of $\langle n \rangle$ and $f$, evolving the system until a NESS is reached. By visual inspection of the simulated trajectories we notice that, for large $\langle n \rangle$, the typical polymer conformations become more crumpled  and display kinks which do not appear in passive ($f=0$) systems (Fig.~\ref{fig:snaps}(a-b)).

To quantify the extent of polymer crumpling we measure the radius of gyration $R_g$ defined by 
\begin{equation}
\label{Rg}
R_g^2 = \left\langle \frac{1}{L}\sum_{i=1}^{L} \left[ \mathbf{r}_i - \mathbf{R}_{\rm cm} \right]^2 \right\rangle ,
\end{equation}
where $\mathbf{r}_i$ is the position of the $i$-th polymer bead, $\mathbf{R}_{\rm cm}$ is the centre of mass of all polymer beads, and $\langle \, \dots \rangle$ denotes ensemble and time average (after the system reaches a steady state). We also measure the persistence length $L_p$ as defined from the bond-bond correlation function
\begin{equation}
\label{Lp}
\langle \mathbf{b}_i\cdot \mathbf{b}_{i+s} \rangle = e^{-s/L_p},  
\end{equation} 
where $\mathbf{b}_{i}=\left( \mathbf{r}_{i+1}-\mathbf{r}_i \right) / |\mathbf{r}_{i+1}-\mathbf{r}_i|$ represents the normalised tangent to the polymer at bead $i$. Here the average is additionally evaluated over bead index $i$. To calculate $L_p$ we perform an exponential fit of the bond-bond correlation function 
(even for cases in which this displays negative values -- see below for more on this point).

As shown in Figs.~\ref{fig:snaps}\textbf{(c)} and \textbf{(d)}, both $R_g$ and $L_p$ exhibit a marked decrease with increasing $\langle n \rangle$, in agreement with our initial qualitative observation. Furthermore, the decrease in effective persistence length $L_p$ is found to have a power-law scaling 
\[
L_p \sim \left( \langle n \rangle f^2 \right)^{\alpha},
\]
with $\alpha\approx -0.1$, which can also be used to collapse the different curves for $R_g$ onto a universal master curve (Fig.~\textbf{(d)} inset). In other words we find that the decrease in average size of the polymer can be well described by a scaling function of the parameter $\langle n \rangle f^2$. 

\begin{figure}[t]
	\centering
	\hspace{-10pt}\includegraphics{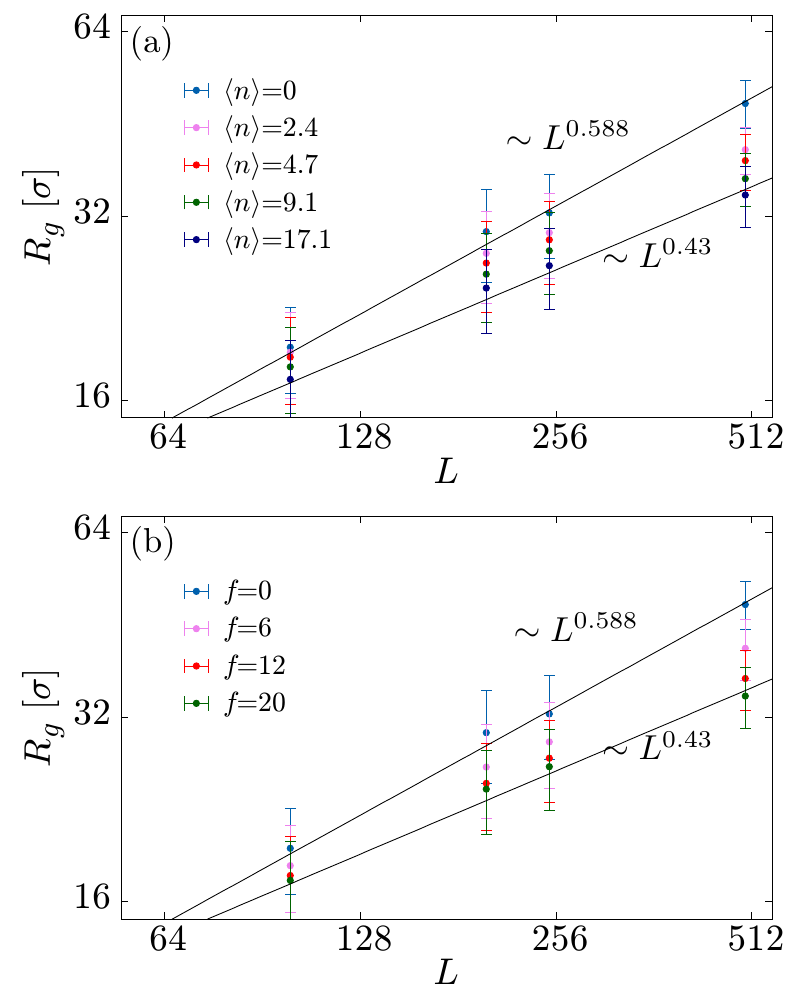}
	\caption{\small Radius of gyration of the polymer substrate as a function of substrate lengths $L$, for \textbf{(a)} a fixed motor force ($f\sigma/k_BT=$20) and different values for the total number of motors $N$; and \textbf{(b)} a fixed number of motors ($N=$200) and different values of the motor force $f$. 
	}
	\label{fig:scaling}
\end{figure}

In order to shed more light onto the effect of the motors on the metric exponent $\nu$ relating the polymer size to its length ($R_g \sim L^{\nu}$), in Fig.~\ref{fig:scaling} we plot $R_g$ a function of $L$ for different combinations of the number of motors $N$ and force $f$. We observe that passive systems ($N=0$ or $f=0$) closely follow the expected self-avoiding statistics ($\nu\approx0.588$). More interestingly, larger values of $N$ or $f$ lead to polymer behaviours that are better fitted by $\nu \simeq$ 0.43. This value lies between that of an ideal chain $\nu = 1/2$ and that of a fully collapsed conformation $\nu=1/3$.  It is intriguing to note that a simple scaling argument can partially explain this finding using the value of $\alpha$ measured in Fig.~\ref{fig:bond-bond-corr} (see Appendix \ref{sec:scaling}).


\begin{figure}[h!]
	\centering
	\includegraphics{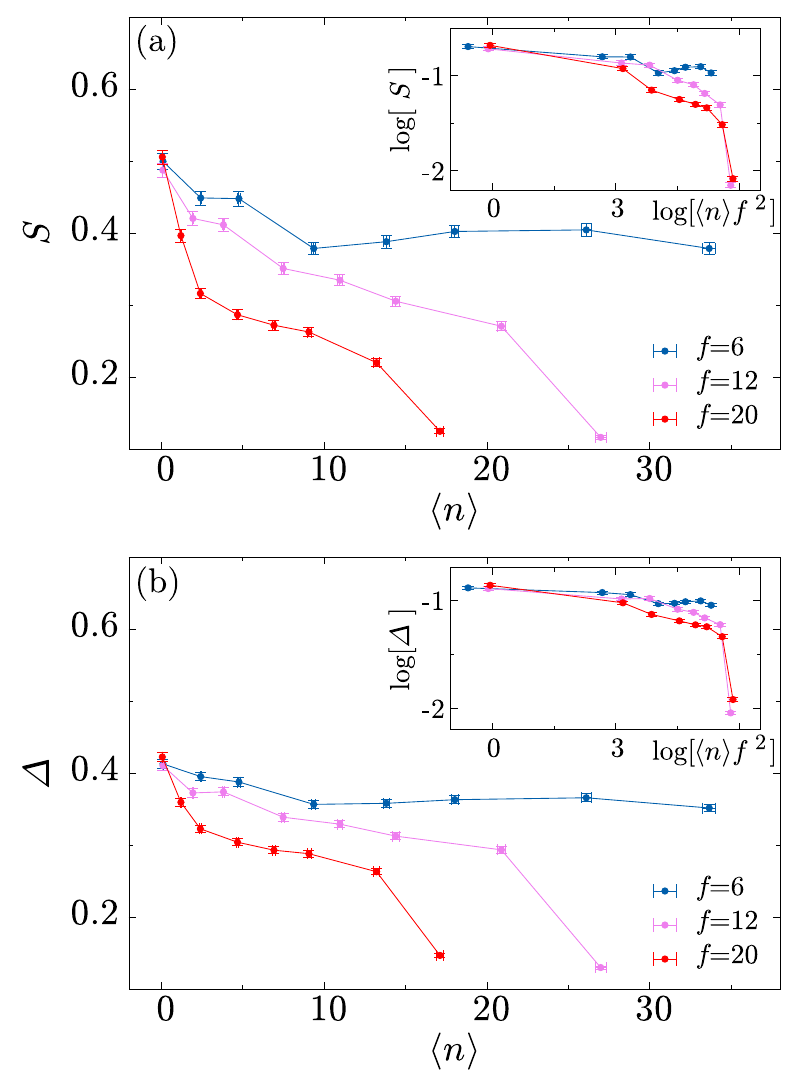}
	\caption{\small \textbf{(a)} Polymer prolateness $S$ as a function of $\langle n \rangle$, for three different values of the motor force $f$. \textbf{(b)} Polymer asphericity $\Delta$ shows a similar trend. Insets show curve collapse when plotted as a function of $\langle n\rangle f^2$ in a double logarithmic scale.}
	\label{fig:aspher}
\end{figure}

To conclude this section we examine the effect of the motors on the global ``shape'' of the polymer by measuring two quantities: the asphericity $\Delta$ and prolateness $S$, which quantify the degree of spherical symmetry and the oblateness of the polymer respectively (see Appendix \ref{sec:shape} for formal definitions). In Fig.~\ref{fig:aspher}, one can readily observe that for increasing $\langle n\rangle$ and $f$ both shape descriptors decrease. This indicates that the coil is becoming more spherical, in agreement with our previous observations that the polymer effectively crumples under the action of the motors. 

\subsection*{Effective softening is explained by hairpin formation}

\begin{figure*}[h!]
	\centering
	\includegraphics{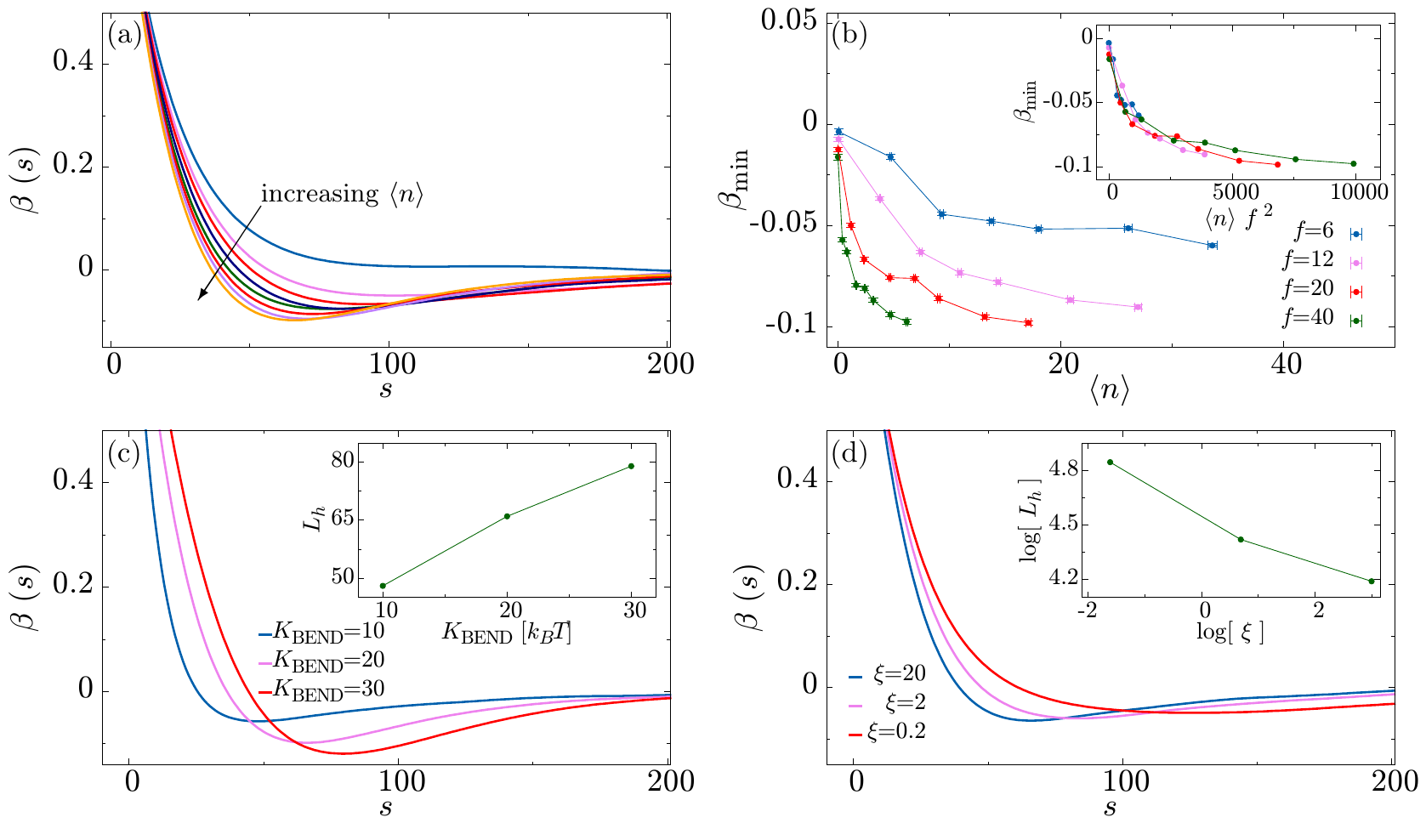}
	\caption{\small\textbf{{(a)}} Polymer bond correlation as a function of the distance along the polymer backbone $s$, at fixed motor force $f = 20$. The arrow indicates the direction of increasing $\langle n \rangle$.  \textbf{(b)}  Plot showing the minimum value of $\beta(s)$ as a function of $\langle n \rangle$. \textbf{Inset} Scaling of the minimum value of this functions, $\beta_{\rm min}$, with $\langle n\rangle f^2$. \textbf{(c)} Polymer bond correlation as a function of the distance along the polymer backbone $s$, at fixed motor force $f = 20$, $\langle n \rangle = 17 $ and different (initial) persistence lengths $L_p$. \textbf{Inset} Position of the bond correlation minimum, $L_{h}$ (defined as the value which satisfies $\beta(s=L_h)=\beta_{\rm min}$), as a function of the polymer persistence length $L_p$. 
	\textbf{(d)} Plot of the polymer bond correlation for the case of motor force $f = 20$ and different values of the friction $\xi$ experienced by the motors. \textbf{Inset} Position of the bond correlation minimum $L_{h}$ as a function of the friction $\xi$.}
	\label{fig:bond-bond-corr}
\end{figure*}

Having established a crumpling effect on the global polymer conformation, we now examine the microscopic mechanism driving this process. We measure the polymer bond auto-correlation function 
\[
\beta(s) \equiv  \left\langle \textbf{b}_i\cdot \textbf{b}_{i+s} \right\rangle,
\]
where as before angle brackets denote the time and ensemble average, and average over $i$.  For an equilibrium polymer $\beta(s)$ decays exponentially to zero (the bond orientation becomes uncorrelated for contour separations much larger than the persistence length). If $\beta(s)$ is close to one for a given $s$ this indicates that bonds with that separation are on average oriented parallel to each other; whereas if $\beta(s)\rightarrow-1$, then bonds with that separation are on average anti-parallel. As shown in Fig.~\ref{fig:bond-bond-corr}(a), $\beta(s)$ drops below zero and displays a minimum -- this indicates a non-negligible probability that bonds separated by about $75$ beads are anti-parallel. In other words, the polymer conformations display hairpins, or ``branches''~\cite{Michieletto2016softmatter} and the distance $s$ for which $\beta(s)$ has a minimum approximately corresponds to the length of the hairpins that are formed on the polymer. The minimum $\beta_{\rm min}$ becomes deeper as both $\langle n \rangle$ and $f$ increase (Fig.~\ref{fig:bond-bond-corr}(b)). In other words, higher values of the motor force cause a more significant anti-correlation of the bonds, i.e. more pronounced hairpins.
A fit of $\beta_{min}$ to the function $f(x) = a x^b$ gives us its scaling as a function of $\langle n \rangle$, with coefficients $a = -0.047 \pm 0.003$ and $b = 0.27 \pm 0.03$. An analogous scaling law can be found for $\beta_{min}$ as a function of the force $f$, this time with coefficients $a = -0.024 \pm  0.006 $ and $b = 0.39 \pm 0.08$.

{A possible physical mechanism underlying hairpin formation is that the motors exert localised tangential forces on the semiflexible polymer they move on. As a result the portion of the polymer ``behind'' the motor (i.e., the polymer region through which the motor has just moved) is under compression, while the portion ahead is under tension -- the situation is reminiscent of forced translocation through a pore~\cite{Saito2013}. The conformation of the polymer could then be computed by minimising its bending energy in the presence of such a localised force. This problem may be viewed as a generalisation of the Euler buckling problem~\cite{LandauLifschitzElasticity,likos_hairpin}. Although a detailed calculation is outside the scope of this work, it is reasonable to expect the combination of tensile and compressive forces to result in local hairpin formation.} 

In Fig.~\ref{fig:bond-bond-corr}(c) we analyse the dependence of the bond correlation function on the intrinsic stiffness of the polymer set by $K_{\rm BEND}$ (recall that in the absence of motor activity, the persistence length $L_p \simeq K_{\rm BEND} \sigma/k_BT$). To this end, we perform three sets of simulations with $K_{\rm BEND} = 10,20$ {and} $30~{k_BT}$, for the case of $\langle n \rangle = 17$ and $f = 20$. Plotting $\beta(s)$ we observe that the \textit{position} of the minimum becomes progressively shifted towards higher values of $s$ as $K_{\rm BEND}$ increases, i.e. the hairpins  become larger (Fig.~\ref{fig:bond-bond-corr}(c) inset). Additionally, we find that the {\textit value} of $\beta(s)$ at the minimum, $\beta_{min}$, gets closer to zero for smaller values of $K_{\rm BEND}$.


In Fig.\ref{fig:bond-bond-corr}(d) we study how changing the relative impact of the active force on the motors and polymer beads affects the hairpin formation. This is achieved by changing the friction experienced by the motors (i.e. we change the viscous drag only for the motors -- this changes their effective hydrodynamic size, or could reflect a difference in interactions with macromolecular crowders compared to the polymer). Increasing the friction for the motors means that they will move less 
as a result of the active force {(which is kept at a constant magnitude)}. Figure~\ref{fig:bond-bond-corr}(d) shows that as the friction for the motors is increased relative to that of the polymer the minimum of $\beta(s)$ shifts to shorter $s$, and gets deeper and narrower. This implies that greater motor friction leads to shorter and tighter hairpins. {We can understand this by considering the opposite limit of very low motor friction; in that case the motors will move more quickly relative to the polymer. Since a motor spends less time at a given polymer bead, that bead will experience the active force for less time and we can expect the effect of the motors to be diminished. The polymer will, however, still experience the active force, so we still expect a ballistic component to its motion.}


\subsection*{Motor activity gives rise to anomalous or enhanced diffusion}

We now turn our attention to the role of molecular motors on the overall polymer dynamics. To this end, we measure the mean square displacement of the polymer centre of mass
\[
{\rm MSD}_{\rm CM}(t) = \left\langle \left[ \mathbf{R}_{\rm cm}(t_0+t) - \mathbf{R}_{\rm cm}(t_0) \right]^2 \right\rangle,
\]
where the average is over start times $t_0$, and $\mathbf{R}_{\rm cm}(t)$ is the position of the centre of mass at time $t$.

\begin{figure}[h!]
	\centering
	\includegraphics[width=.5\textwidth]{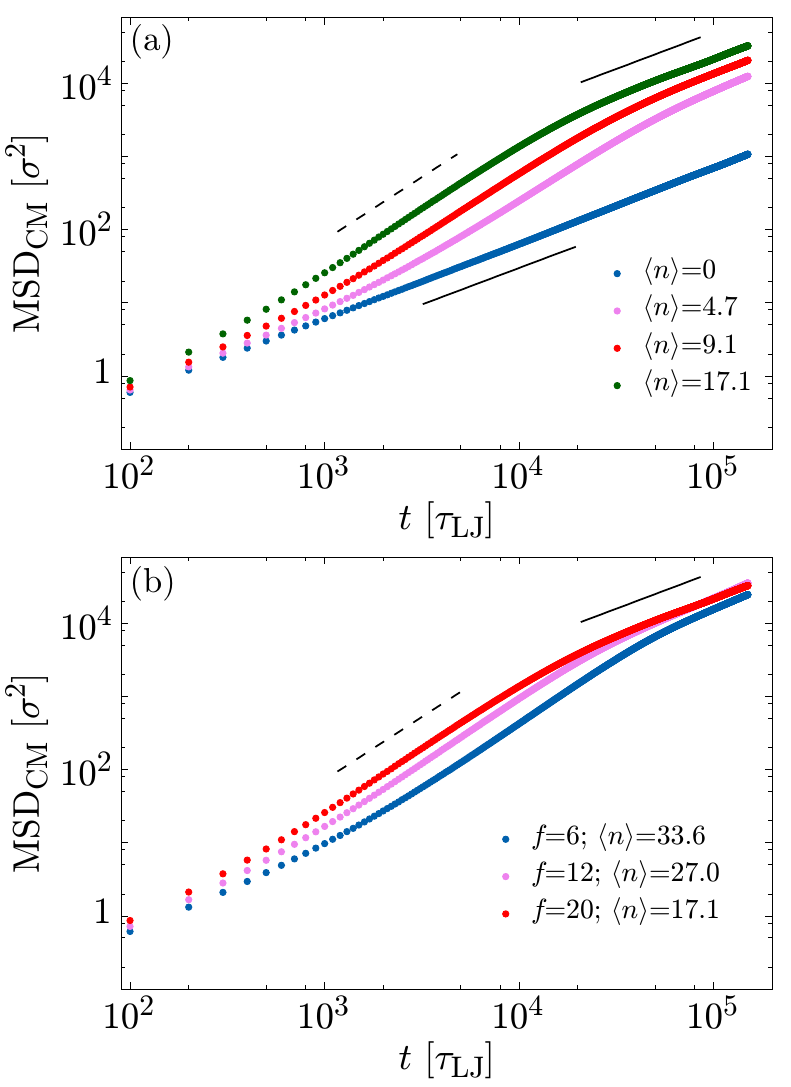}
	\caption{\small \textbf{(a)} Mean squared displacement of the polymer centre of mass as a function of time. Each set of points shows results for a different value of $\langle n \rangle$ at fixed $f=20~k_BT \sigma^{-1}$. Higher values of $\langle n \rangle$ induce a faster diffusivity of the polymer.  \textbf{(b)} Similar plot but now each set of points shows results for a different value of $f$, with the total number of motors fixed at $N=400$. 
}
	\label{fig:Fig_diff}
\end{figure}

Figure~\ref{fig:Fig_diff} shows that at short times the system is diffusive as the propulsion exerted by the motors needs a characteristic time in order to provide a displacement comparable to a Brownian displacement. Following the Rouse model~\cite{rouse} ${\rm MSD}_{\rm CM}(t) = (6D_0/L) t$, where $D_0 = k_B T/\xi$ is the diffusion coefficient of a single monomer. This behaviour is recovered at all times for $\langle n \rangle \rightarrow 0$.  

\begin{table}[h!]
	\begin{center}
		\begin{tabular}{ |c|c|c| } 
			\hline
			\multicolumn{3}{|c|}{\textbf{Intermediate times: super-diffusive}}\\
			\hline
			$\langle n \rangle$ & $\alpha$ & $\mu$ \\
			\hline
			17 & $1.720\pm 0.002$ & $3.1\cdot10^{-5}\pm0.5\cdot10^{-6}$ \\ 
			9 & $1.690 \pm 0.001$ & $1.7\cdot10^{-5} \pm 0.2\cdot10^{-6}$ \\ 
			5 & $1.542 \pm 0.006 $ & $2.6\cdot10^{-5} \pm 1.3\cdot10^{-6}$\\
			\hline
			\multicolumn{3}{|c|}{\textbf{Long times: diffusive}}\\
			\hline
			$\langle n \rangle$ & $\alpha$ & $D$ \\
			\hline
			17 & $1.025 \pm0.001 $& $2.7\cdot10^{-2} \pm 2.6\cdot 10^{-4}$ \\
			9 & $1.020 \pm 0.002 $ & $2.0\cdot10^{-2} \pm 0.0 6\cdot10^{-4}$\\
			5 & $1.090 \pm 0.0001$& $0.4\cdot10^{-2} \pm 0.08\cdot10^{-4}$\\
			\hline
		\end{tabular}
	\end{center} 
	\caption{Diffusion properties of the centre of mass of the polymer are measured by fitting to the MSD for different numbers of motors. Fits are evaluated using the function $f(t) = a\cdot x^{\alpha}$, and the diffusion coefficient (or mobility $\mu$) is then $D = a/6$ following the Rouse model.}
	\label{table:msd}
\end{table}

At intermediate times, the polymer exhibits super-diffusive motion, i.e. ${\rm MSD}_{\rm CM}(t) \sim t^{\alpha}$ with $\alpha \sim 1.6 $.  Finally, at large times we observed enhanced diffusion, i.e. $\alpha\approx1$ but the diffusion constant increases with motor activity ($f$ or $\langle n \rangle$).  In Table~\ref{table:msd} we report the measured diffusion constant $D$ for the different cases considered. 
One can readily observe that the exponent $\alpha$ is roughly independent of $\langle n \rangle$ and $f$, whereas the diffusion coefficient increases with $\langle n \rangle$.  Moreover, by comparing fits at short and long times for each value of $\langle n \rangle$, we find that the crossover between the super-diffusive and diffusive behaviour occurs at progressively larger times as $\langle n \rangle$ decreases.
Together, this implies that the motor action leads to a persistent motion of the centre of mass of the polymer; at longer times the velocity becomes uncorrelated and the motion is again diffusive. 

{These three regimes are typical of other active systems, e.g. in polymers where an active force directed along the backbone is applied to each polymer bead~\cite{bianco_loca,Anand2018} and in  Active Brownian Particles (ABPs)~\cite{Cates2012}. At intermediate times, the motion of an ABP is dictated by a force which typically maintains its direction for a characteristic time, leading to super-diffusion. In our model, a correlation on the scale of the whole polymer is set by the end-to-end vector.}


\subsection*{Untying Knots with Molecular Motors}

Finally, we discuss a possible application of our model by studying the knotting probability of a polymer in non-equilibrium conditions.  We start from the observation that a freely diffusing linear polymer in a good solvent may display a physical knot -- i.e. a knot within an open segment of the polymer contour -- with probability $P_{\rm knot}$. It has recently been shown~\cite{matthews2012, poier2014, Coronel2017} that $P_{\rm knot}$ displays a non-monotonic dependence on the stiffness of the substrate with a maximum at around $L_p \simeq 5-10 \sigma$. In light of the results presented above -- i.e., motors drive a decrease in the apparent persistence length of the substrate -- we reason that the presence of the motors should lead to a shift in the knotting probability as a function of substrate persistence length. Specifically, we expect that, due to the action of the motors, the maximum in $P_{\rm knot}$ will be located at larger values of the intrinsic persistence length of the substrate, $l_p$ (to be clear, the lower case $l_p$ refers to the persistence length the polymer would have in the absence of motors).

\begin{figure}[t!]
	\centering
	\includegraphics{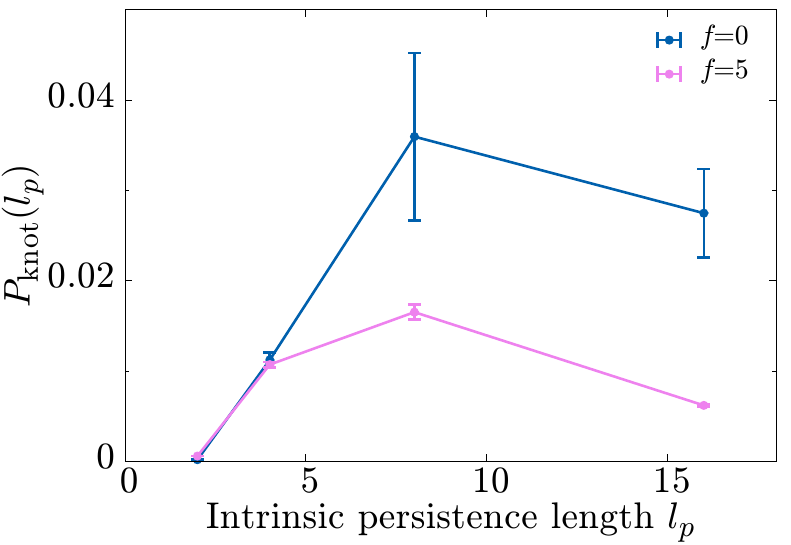}
	\vspace{-0,35 cm}
	\caption{Molecular motors acting on a polymeric substrate reduce the knotting probability $P_{\rm knot}(l_p)$ of a free chain in a good solvent. Here $l_p$ is the intrinsic persistence length which would be measured in the absence of motors. The simulations are performed on a 500 bead long chain and the points are averages over $10^4$ independent configurations (error bars show the standard error in the mean).} 
	\label{fig:knots}
\end{figure}

In order to test this hypothesis we perform simulations of a freely diffusing linear polymer 500 beads long, and measure the probability of forming physical knots as a function of $l_p$ by adopting the algorithm described in Refs.~\cite{Tubiana2011,Tubiana2011prl} and publicly available at \url{http://kymoknot.sissa.it/}~\cite{Tubiana2018}. By comparing $P_{\rm knot}(l_p)$ for cases with and without motors we find that, surprisingly, the location of the maximum of $P_{\rm knot}$ is unchanged by the action of the motors, i.e. $l_p \simeq 5-10 \sigma$. The most apparent effect of the motor activity is a global reduction in the knotting probability across the full spectrum of bending rigidities studied (Fig.~\ref{fig:knots}). 
In light of this finding we argue that the effect of the motors goes beyond a simple softening of the polymer backbone, and that subtler non-equilibrium effects must be at play. 
We also speculate that the reduced knotting probability due to motor activity may be relevant for the case of DNA under the action of motors such as polymerase. According to our model, the action of these motors would naturally have the beneficial consequence of reducing the occurrence of knots and self-entanglements {\textit in vivo}~\cite{Bates2005}.

\section{Conclusions}

In this work we performed Langevin dynamics simulations of a coarse grained model to study the effect of molecular motors on a polymer substrate. We examined both the static and dynamic properties of polymers in dilute conditions while varying the number of motors and the magnitude of the force which they generate.

Our main result was that the effect of the motors is to reduce the overall size of the polymer coil by increasing the likelihood of double-folded segments, or hairpins. This ``softening'' of the backbone manifests as a reduction of the measured persistence length of the polymer. We found that the data collapse onto a single curve when plotted as a function of $\langle n\rangle f^2$, revealing a power-law scaling $R_g \sim (\langle n\rangle f^2)^{-0.1}$ and $L_p \sim (\langle n\rangle f^2)^{-0.1}$ (insets of Fig.~\ref{fig:snaps}\textbf{(c)} and \textbf{(d)}).  We also performed simulations of polymers with different lengths and obtained the explicit values of the metric exponent $\nu$ (where $R_g \sim L^{\nu}$) as a function of $\langle n\rangle$ and $f$. These confirm that at equilibrium the polymers obey self-avoiding statistics, whereas motor activity leads to an exponent between those expected for a crumpled globule and an ideal chain (Fig.~\ref{fig:scaling}).
{A reduction in $R_g$ and $L_p$ has also been observed in models without explicit motors, where an active force directed tangentially to the polymer is added directly to each polymer bead~\cite{bianco_loca,Anand2018}. In that case the softening arises because, for an initially straight segment of the polymer (e.g. of length equal to the persistence length), small fluctuations in backbone direction will lead to components of the tangential force which are perpendicular to the end-to-end vector for the segment; this then drives the polymer into large smooth curves~\cite{bianco_loca}. Those studies, however, did not find hairpins (the decay in bond correlations remained exponential~\cite{Anand2018}), suggesting the present case involves a different mechanism. }

We {also} found that the motors affect the dynamics of the polymer. At intermediate times the centre of mass displays super-diffusive behaviour while at longer times it follows an enhanced diffusion. Both regimes are controlled by motor activity (number of bound motors and force applied).  While a super-diffusive regime was to be expected, our simulation have provided a detailed quantification of how that can be tuned through the action of the motors (Fig.~\ref{fig:Fig_diff} and Table~\ref{table:msd}). 
The effect of the motors on the knotting probability of the polymer was subtler. Unexpectedly, $P_{\rm knot}(l_p)$ did not simply show a shift due to the change in effective persistence length, but rather there was a global reduction in the steady state knotting probability across a range of substrate rigidities (Fig.~\ref{fig:knots}). {We can speculate that since the motor action promotes polymer configurations with hairpin bends, but suppresses knotted ones, then these two types of configuration are incompatible. An interesting future study would be to examine knotting probabilities in other active polymer models where hairpins are not observed.}

Throughout this paper we have limited ourselves to the simple case in which all of the motors travel in the same direction along the substrate.  While this is the case for some systems, e.g. kinesin on microtubules, in others the motors can travel in either direction.  For example, RNA polymerase follows the direction of genes, depending on the specifically oriented DNA sequences in gene promoters. This could be readily incorporated into our model in the future by either considering a mixture of two species of motors which move in different directions on the substrate, or by setting some directional binding sites. {Also, we have considered the case of motors which are the same size as the polymer beads -- in reality RNA polymerase is a large molecule of around 30~nm in size. However, as our results are mostly affected by the magnitude of the force, we do not expect motor size to have a qualitative affect. Nevertheless it would be interesting to examine systems with larger motors in the future.}

Another interesting scenario would be to consider a mixture of passive and active polymers. Indeed, an activity-driven phase separation has been observed in another non-equilibrium polymer system -- where two species of polymer were held at different temperatures~\cite{Smrek2017}. It would be interesting to see whether such demixing behaviour can be recovered when the system is driven away from equilibrium by a more biologically relevant mechanism, such as that presented here. 

Our model naturally lends itself to the study of the response of polymer substrates to the action of molecular motors. We therefore expect that in the future is could be employed to address more specific biological questions, such as understanding the role of motors in the buckling of actin networks~\cite{Broedersz2014,Lenz2014}, or in the bending and active collective motion of microtubules~\cite{Head2011,Pearce2018} by suitably tuning the rigidity of the substrate and processivity of the motors.  

Finally we stress that the model presented here is among the first to explicitly account for the presence of physical motors; as a consequence, it faithfully captures the feedback between the action of the motors and the change in substrate-motor interaction due to, for instance, alterations in local polymer conformation. For this reason, we believe that our model will be pivotal in studying complex active collective phenomena where interactions between motors are mediated through the target substrate.

\subsection*{Acknowledgements}

This research was partly funded by the European Training Network COLLDENSE (N2020-MCSA-ITN-2014) grant number 642774.
DMa, DMi and CAB acknowledge funding from ERC (CoG 648050 THREEDCELLPHYSICS). \\
DMi, CNL and EL would also like to acknowledge the contribution and networking support by the ``European Topology Interdisciplinary Action'' (EUTOPIA) CA17139.
\footnotesize{

\begin{mcitethebibliography}{47}
\providecommand*{\natexlab}[1]{#1}
\providecommand*{\mciteSetBstSublistMode}[1]{}
\providecommand*{\mciteSetBstMaxWidthForm}[2]{}
\providecommand*{\mciteBstWouldAddEndPuncttrue}
  {\def\EndOfBibitem{\unskip.}}
\providecommand*{\mciteBstWouldAddEndPunctfalse}
  {\let\EndOfBibitem\relax}
\providecommand*{\mciteSetBstMidEndSepPunct}[3]{}
\providecommand*{\mciteSetBstSublistLabelBeginEnd}[3]{}
\providecommand*{\EndOfBibitem}{}
\mciteSetBstSublistMode{f}
\mciteSetBstMaxWidthForm{subitem}
{(\emph{\alph{mcitesubitemcount}})}
\mciteSetBstSublistLabelBeginEnd{\mcitemaxwidthsubitemform\space}
{\relax}{\relax}

\bibitem[Sanchez \emph{et~al.}(2012)Sanchez, Chen, Decamp, Heymann, and
  Dogic]{Sanchez2012nature}
T.~Sanchez, D.~T.~N. Chen, S.~J. Decamp, M.~Heymann and Z.~Dogic,
  \emph{Nature}, 2012, \textbf{491}, 1--5\relax
\mciteBstWouldAddEndPuncttrue
\mciteSetBstMidEndSepPunct{\mcitedefaultmidpunct}
{\mcitedefaultendpunct}{\mcitedefaultseppunct}\relax
\EndOfBibitem
\bibitem[Alberts \emph{et~al.}(2014)Alberts, Johnson, Lewis, Morgan, and
  Raff]{Alberts2014}
B.~Alberts, A.~Johnson, J.~Lewis, D.~Morgan and M.~Raff, \emph{{Molecular
  Biology of the Cell}}, Taylor {\&} Francis, 2014, p. 1464\relax
\mciteBstWouldAddEndPuncttrue
\mciteSetBstMidEndSepPunct{\mcitedefaultmidpunct}
{\mcitedefaultendpunct}{\mcitedefaultseppunct}\relax
\EndOfBibitem
\bibitem[Calladine \emph{et~al.}(1997)Calladine, Drew, Luisi, Travers, and
  Bash]{Calladine1997}
C.~R. Calladine, H.~Drew, F.~B. Luisi, A.~A. Travers and E.~Bash,
  \emph{{Understanding DNA: the molecule and how it works}}, Elsevier Academic
  Press, 1997, vol.~1\relax
\mciteBstWouldAddEndPuncttrue
\mciteSetBstMidEndSepPunct{\mcitedefaultmidpunct}
{\mcitedefaultendpunct}{\mcitedefaultseppunct}\relax
\EndOfBibitem
\bibitem[Le \emph{et~al.}(2013)Le, Imakaev, Mirny, and Laub]{Le2013}
T.~B. Le, M.~V. Imakaev, L.~A. Mirny and M.~T. Laub, \emph{Science (New York,
  N.Y.)}, 2013, \textbf{342 VN -}, 731--734\relax
\mciteBstWouldAddEndPuncttrue
\mciteSetBstMidEndSepPunct{\mcitedefaultmidpunct}
{\mcitedefaultendpunct}{\mcitedefaultseppunct}\relax
\EndOfBibitem
\bibitem[Terakawa \emph{et~al.}(2017)Terakawa, Bisht, Eeftens, Dekker, Haering,
  and Greene]{Terakawa2017}
T.~Terakawa, S.~Bisht, J.~M. Eeftens, C.~Dekker, C.~H. Haering and E.~C.
  Greene, \emph{Science}, 2017, \textbf{676}, eaan6516\relax
\mciteBstWouldAddEndPuncttrue
\mciteSetBstMidEndSepPunct{\mcitedefaultmidpunct}
{\mcitedefaultendpunct}{\mcitedefaultseppunct}\relax
\EndOfBibitem
\bibitem[Brackley \emph{et~al.}(2016)Brackley, Johnson, Bentivoglio, Corless,
  Gilbert, Gonnella, and Marenduzzo]{Brackley2016supercoil}
C.~A. Brackley, J.~Johnson, A.~Bentivoglio, S.~Corless, N.~Gilbert, G.~Gonnella
  and D.~Marenduzzo, \emph{Phys. Rev. Lett.}, 2016, \textbf{117}, 018101\relax
\mciteBstWouldAddEndPuncttrue
\mciteSetBstMidEndSepPunct{\mcitedefaultmidpunct}
{\mcitedefaultendpunct}{\mcitedefaultseppunct}\relax
\EndOfBibitem
\bibitem[Naughton \emph{et~al.}(2013)Naughton, Avlonitis, Corless, Prendergast,
  Mati, Eijk, Cockroft, Bradley, Ylstra, and Gilbert]{Naughton2013}
C.~Naughton, N.~Avlonitis, S.~Corless, J.~G. Prendergast, I.~K. Mati, P.~P.
  Eijk, S.~L. Cockroft, M.~Bradley, B.~Ylstra and N.~Gilbert, \emph{Nat.
  Struct. Mol. Biol.}, 2013, \textbf{20}, 387--395\relax
\mciteBstWouldAddEndPuncttrue
\mciteSetBstMidEndSepPunct{\mcitedefaultmidpunct}
{\mcitedefaultendpunct}{\mcitedefaultseppunct}\relax
\EndOfBibitem
\bibitem[Lenz(2014)]{Lenz2014}
M.~Lenz, \emph{Phys. Rev. X}, 2014, \textbf{4}, 1--9\relax
\mciteBstWouldAddEndPuncttrue
\mciteSetBstMidEndSepPunct{\mcitedefaultmidpunct}
{\mcitedefaultendpunct}{\mcitedefaultseppunct}\relax
\EndOfBibitem
\bibitem[Broedersz and MacKintosh(2014)]{Broedersz2014}
C.~P. Broedersz and F.~C. MacKintosh, \emph{Rev. Mod. Phys.}, 2014,
  \textbf{86}, 995--1036\relax
\mciteBstWouldAddEndPuncttrue
\mciteSetBstMidEndSepPunct{\mcitedefaultmidpunct}
{\mcitedefaultendpunct}{\mcitedefaultseppunct}\relax
\EndOfBibitem
\bibitem[Tjhung \emph{et~al.}(2012)Tjhung, Marenduzzo, and Cates]{Tjhung2012}
E.~Tjhung, D.~Marenduzzo and M.~E. Cates, \emph{Proc. Natl. Acad. Sci. USA},
  2012, \textbf{109}, 12381--12386\relax
\mciteBstWouldAddEndPuncttrue
\mciteSetBstMidEndSepPunct{\mcitedefaultmidpunct}
{\mcitedefaultendpunct}{\mcitedefaultseppunct}\relax
\EndOfBibitem
\bibitem[Herbert \emph{et~al.}(2008)Herbert, Greenleaf, and Block]{Herbert2008}
K.~M. Herbert, W.~J. Greenleaf and S.~M. Block, \emph{Annu Rev Biochem}, 2008,
  \textbf{77}, 149--176\relax
\mciteBstWouldAddEndPuncttrue
\mciteSetBstMidEndSepPunct{\mcitedefaultmidpunct}
{\mcitedefaultendpunct}{\mcitedefaultseppunct}\relax
\EndOfBibitem
\bibitem[Michaelis and Treutlein(2013)]{michaelis2013single}
J.~Michaelis and B.~Treutlein, \emph{Chemical reviews}, 2013, \textbf{113},
  8377--8399\relax
\mciteBstWouldAddEndPuncttrue
\mciteSetBstMidEndSepPunct{\mcitedefaultmidpunct}
{\mcitedefaultendpunct}{\mcitedefaultseppunct}\relax
\EndOfBibitem
\bibitem[Ganji \emph{et~al.}(2018)Ganji, Shaltiel, Bisht, Kim, Kalichava,
  Haering, and Dekker]{Ganji2018}
A.~M. Ganji, I.~A. Shaltiel, S.~Bisht, E.~Kim, A.~Kalichava, C.~H. Haering and
  C.~Dekker, \emph{Science}, 2018, \textbf{7831}, 1--9\relax
\mciteBstWouldAddEndPuncttrue
\mciteSetBstMidEndSepPunct{\mcitedefaultmidpunct}
{\mcitedefaultendpunct}{\mcitedefaultseppunct}\relax
\EndOfBibitem
\bibitem[Bianco \emph{et~al.}(2018)Bianco, Locatelli, and
  Malgaretti]{bianco_loca}
V.~Bianco, E.~Locatelli and P.~Malgaretti, \emph{Phys. Rev. Lett.}, 2018,
  \textbf{121}, 217802\relax
\mciteBstWouldAddEndPuncttrue
\mciteSetBstMidEndSepPunct{\mcitedefaultmidpunct}
{\mcitedefaultendpunct}{\mcitedefaultseppunct}\relax
\EndOfBibitem
\bibitem[Saintillan \emph{et~al.}(2018)Saintillan, Shelley, and
  Zidovska]{SaintillanMotor}
D.~Saintillan, M.~J. Shelley and A.~Zidovska, \emph{Proceedings of the National
  Academy of Sciences}, 2018, \textbf{115}, 11442--11447\relax
\mciteBstWouldAddEndPuncttrue
\mciteSetBstMidEndSepPunct{\mcitedefaultmidpunct}
{\mcitedefaultendpunct}{\mcitedefaultseppunct}\relax
\EndOfBibitem
\bibitem[Harder \emph{et~al.}(2014)Harder, Valeriani, and Cacciuto]{Harder2014}
J.~Harder, C.~Valeriani and A.~Cacciuto, \emph{Phys. Rev. E}, 2014,
  \textbf{90}, 062312\relax
\mciteBstWouldAddEndPuncttrue
\mciteSetBstMidEndSepPunct{\mcitedefaultmidpunct}
{\mcitedefaultendpunct}{\mcitedefaultseppunct}\relax
\EndOfBibitem
\bibitem[Jiang and Hou(2014)]{Jiang2014}
H.~Jiang and Z.~Hou, \emph{Soft Matter}, 2014, \textbf{10}, 1012--1017\relax
\mciteBstWouldAddEndPuncttrue
\mciteSetBstMidEndSepPunct{\mcitedefaultmidpunct}
{\mcitedefaultendpunct}{\mcitedefaultseppunct}\relax
\EndOfBibitem
\bibitem[Isele-Holder \emph{et~al.}(2015)Isele-Holder, Elgeti, and
  Gompper]{Isele-Holder2015}
R.~E. Isele-Holder, J.~Elgeti and G.~Gompper, \emph{Soft Matter}, 2015,
  \textbf{11}, 7181--7190\relax
\mciteBstWouldAddEndPuncttrue
\mciteSetBstMidEndSepPunct{\mcitedefaultmidpunct}
{\mcitedefaultendpunct}{\mcitedefaultseppunct}\relax
\EndOfBibitem
\bibitem[Eisenstecken \emph{et~al.}(2016)Eisenstecken, Gompper, and
  Winkler]{Eisenstecken2016}
T.~Eisenstecken, G.~Gompper and R.~G. Winkler, \emph{Polymers}, 2016,
  \textbf{8}, year\relax
\mciteBstWouldAddEndPuncttrue
\mciteSetBstMidEndSepPunct{\mcitedefaultmidpunct}
{\mcitedefaultendpunct}{\mcitedefaultseppunct}\relax
\EndOfBibitem
\bibitem[Anand and Singh(2018)]{Anand2018}
S.~K. Anand and S.~P. Singh, \emph{Phys. Rev. E}, 2018, \textbf{98},
  042501\relax
\mciteBstWouldAddEndPuncttrue
\mciteSetBstMidEndSepPunct{\mcitedefaultmidpunct}
{\mcitedefaultendpunct}{\mcitedefaultseppunct}\relax
\EndOfBibitem
\bibitem[Winkler \emph{et~al.}(2017)Winkler, Elgeti, and Gompper]{Winkler2017}
R.~G. Winkler, J.~Elgeti and G.~Gompper, \emph{Journal of the Physical Society
  of Japan}, 2017, \textbf{86}, 101014\relax
\mciteBstWouldAddEndPuncttrue
\mciteSetBstMidEndSepPunct{\mcitedefaultmidpunct}
{\mcitedefaultendpunct}{\mcitedefaultseppunct}\relax
\EndOfBibitem
\bibitem[Schaller \emph{et~al.}(2010)Schaller, Weber, Semmrich, Frey, and
  Bausch]{Schaller2010}
V.~Schaller, C.~Weber, C.~Semmrich, E.~Frey and A.~R. Bausch, \emph{Nature},
  2010, \textbf{467}, 73\relax
\mciteBstWouldAddEndPuncttrue
\mciteSetBstMidEndSepPunct{\mcitedefaultmidpunct}
{\mcitedefaultendpunct}{\mcitedefaultseppunct}\relax
\EndOfBibitem
\bibitem[Sanchez \emph{et~al.}(2012)Sanchez, Chen, Decamp, Heymann, and
  Dogic]{Sanchez2012}
T.~Sanchez, D.~T.~N. Chen, S.~J. Decamp, M.~Heymann and Z.~Dogic,
  \emph{Nature}, 2012, \textbf{491}, 1--5\relax
\mciteBstWouldAddEndPuncttrue
\mciteSetBstMidEndSepPunct{\mcitedefaultmidpunct}
{\mcitedefaultendpunct}{\mcitedefaultseppunct}\relax
\EndOfBibitem
\bibitem[Nedelec and Foethke(2007)]{nedelec2007}
F.~Nedelec and D.~Foethke, \emph{New Journal of Physics}, 2007, \textbf{9},
  427\relax
\mciteBstWouldAddEndPuncttrue
\mciteSetBstMidEndSepPunct{\mcitedefaultmidpunct}
{\mcitedefaultendpunct}{\mcitedefaultseppunct}\relax
\EndOfBibitem
\bibitem[Freedman \emph{et~al.}(2017)Freedman, Banerjee, Hocky, and
  Dinner]{Freedman2017}
S.~L. Freedman, S.~Banerjee, G.~M. Hocky and A.~R. Dinner, \emph{Biophysical
  Journal}, 2017, \textbf{113}, 448--460\relax
\mciteBstWouldAddEndPuncttrue
\mciteSetBstMidEndSepPunct{\mcitedefaultmidpunct}
{\mcitedefaultendpunct}{\mcitedefaultseppunct}\relax
\EndOfBibitem
\bibitem[Ravichandran \emph{et~al.}(2017)Ravichandran, Vliegenthart,
  Saggiorato, Auth, and Gompper]{Ravichandran2017}
A.~Ravichandran, G.~A. Vliegenthart, G.~Saggiorato, T.~Auth and G.~Gompper,
  \emph{Biophysical Journal}, 2017, \textbf{113}, 1121 -- 1132\relax
\mciteBstWouldAddEndPuncttrue
\mciteSetBstMidEndSepPunct{\mcitedefaultmidpunct}
{\mcitedefaultendpunct}{\mcitedefaultseppunct}\relax
\EndOfBibitem
\bibitem[Gupta \emph{et~al.}(2019)Gupta, Chaudhuri, and Chaudhuri]{Gupta2019}
N.~Gupta, A.~Chaudhuri and D.~Chaudhuri, \emph{Phys. Rev. E}, 2019,
  \textbf{99}, 042405\relax
\mciteBstWouldAddEndPuncttrue
\mciteSetBstMidEndSepPunct{\mcitedefaultmidpunct}
{\mcitedefaultendpunct}{\mcitedefaultseppunct}\relax
\EndOfBibitem
\bibitem[Coronel \emph{et~al.}(2017)Coronel, Orlandini, and
  Micheletti]{Coronel2017}
L.~Coronel, E.~Orlandini and C.~Micheletti, \emph{Soft Matter}, 2017,
  \textbf{13}, 4260--4267\relax
\mciteBstWouldAddEndPuncttrue
\mciteSetBstMidEndSepPunct{\mcitedefaultmidpunct}
{\mcitedefaultendpunct}{\mcitedefaultseppunct}\relax
\EndOfBibitem
\bibitem[Plimpton(1995)]{LAMMPS}
S.~Plimpton, \emph{J Comp Phys}, 1995, \textbf{117}, 1--19\relax
\mciteBstWouldAddEndPuncttrue
\mciteSetBstMidEndSepPunct{\mcitedefaultmidpunct}
{\mcitedefaultendpunct}{\mcitedefaultseppunct}\relax
\EndOfBibitem
\bibitem[Michieletto(2016)]{Michieletto2016softmatter}
D.~Michieletto, \emph{Soft Matter}, 2016, \textbf{12}, 9485--9500\relax
\mciteBstWouldAddEndPuncttrue
\mciteSetBstMidEndSepPunct{\mcitedefaultmidpunct}
{\mcitedefaultendpunct}{\mcitedefaultseppunct}\relax
\EndOfBibitem
\bibitem[Saito and Sakaue(2013)]{Saito2013}
T.~Saito and T.~Sakaue, \emph{Phys. Rev. E}, 2013, \textbf{88}, 042606\relax
\mciteBstWouldAddEndPuncttrue
\mciteSetBstMidEndSepPunct{\mcitedefaultmidpunct}
{\mcitedefaultendpunct}{\mcitedefaultseppunct}\relax
\EndOfBibitem
\bibitem[Landau \emph{et~al.}(1986)Landau, Kosevich, Pitaevskii, and
  Lifshitz]{LandauLifschitzElasticity}
L.~D. Landau, A.~Kosevich, L.~Pitaevskii and E.~Lifshitz, \emph{Theory of
  Elasticity}, Butterworth-Heinemann, Oxford, 1986\relax
\mciteBstWouldAddEndPuncttrue
\mciteSetBstMidEndSepPunct{\mcitedefaultmidpunct}
{\mcitedefaultendpunct}{\mcitedefaultseppunct}\relax
\EndOfBibitem
\bibitem[Wynveen and Likos(2010)]{likos_hairpin}
A.~Wynveen and C.~N. Likos, \emph{Soft Matter}, 2010, \textbf{6},
  163--171\relax
\mciteBstWouldAddEndPuncttrue
\mciteSetBstMidEndSepPunct{\mcitedefaultmidpunct}
{\mcitedefaultendpunct}{\mcitedefaultseppunct}\relax
\EndOfBibitem
\bibitem[Prince E.~Rouse()]{rouse}
J.~Prince E.~Rouse, \emph{The Journal of Chemical Physics}, \textbf{20},
  year\relax
\mciteBstWouldAddEndPuncttrue
\mciteSetBstMidEndSepPunct{\mcitedefaultmidpunct}
{\mcitedefaultendpunct}{\mcitedefaultseppunct}\relax
\EndOfBibitem
\bibitem[Cates(2012)]{Cates2012}
M.~E. Cates, \emph{Rep. Prog. Phys.}, 2012, \textbf{75}, 042601\relax
\mciteBstWouldAddEndPuncttrue
\mciteSetBstMidEndSepPunct{\mcitedefaultmidpunct}
{\mcitedefaultendpunct}{\mcitedefaultseppunct}\relax
\EndOfBibitem
\bibitem[Matthews \emph{et~al.}(2012)Matthews, Louis, and Likos]{matthews2012}
R.~Matthews, A.~a. Louis and C.~N. Likos, \emph{ACS Macro Letters}, 2012,
  1352--1356\relax
\mciteBstWouldAddEndPuncttrue
\mciteSetBstMidEndSepPunct{\mcitedefaultmidpunct}
{\mcitedefaultendpunct}{\mcitedefaultseppunct}\relax
\EndOfBibitem
\bibitem[Poier \emph{et~al.}(2014)Poier, Likos, and Matthews]{poier2014}
P.~Poier, C.~N. Likos and R.~Matthews, \emph{Macromolecules}, 2014,
  \textbf{47}, 3394--3400\relax
\mciteBstWouldAddEndPuncttrue
\mciteSetBstMidEndSepPunct{\mcitedefaultmidpunct}
{\mcitedefaultendpunct}{\mcitedefaultseppunct}\relax
\EndOfBibitem
\bibitem[Tubiana \emph{et~al.}(2011)Tubiana, Orlandini, and
  Micheletti]{Tubiana2011}
L.~Tubiana, E.~Orlandini and C.~Micheletti, \emph{Prog. Theor. Phys. Suppl.},
  2011, \textbf{191}, 192--204\relax
\mciteBstWouldAddEndPuncttrue
\mciteSetBstMidEndSepPunct{\mcitedefaultmidpunct}
{\mcitedefaultendpunct}{\mcitedefaultseppunct}\relax
\EndOfBibitem
\bibitem[Tubiana \emph{et~al.}(2011)Tubiana, Orlandini, and
  Micheletti]{Tubiana2011prl}
L.~Tubiana, E.~Orlandini and C.~Micheletti, \emph{Phys. Rev. Lett.}, 2011,
  \textbf{107}, 188302\relax
\mciteBstWouldAddEndPuncttrue
\mciteSetBstMidEndSepPunct{\mcitedefaultmidpunct}
{\mcitedefaultendpunct}{\mcitedefaultseppunct}\relax
\EndOfBibitem
\bibitem[Tubiana \emph{et~al.}(2018)Tubiana, Polles, Orlandini, and
  Micheletti]{Tubiana2018}
L.~Tubiana, G.~Polles, E.~Orlandini and C.~Micheletti, \emph{EPJ E}, 2018,
  \textbf{41}, 72\relax
\mciteBstWouldAddEndPuncttrue
\mciteSetBstMidEndSepPunct{\mcitedefaultmidpunct}
{\mcitedefaultendpunct}{\mcitedefaultseppunct}\relax
\EndOfBibitem
\bibitem[Bates and Maxwell(2005)]{Bates2005}
A.~Bates and A.~Maxwell, \emph{{DNA topology}}, Oxford University Press,
  2005\relax
\mciteBstWouldAddEndPuncttrue
\mciteSetBstMidEndSepPunct{\mcitedefaultmidpunct}
{\mcitedefaultendpunct}{\mcitedefaultseppunct}\relax
\EndOfBibitem
\bibitem[Smrek and Kremer(2017)]{Smrek2017}
J.~Smrek and K.~Kremer, \emph{Phys. Rev. Lett.}, 2017, \textbf{118}, 1--5\relax
\mciteBstWouldAddEndPuncttrue
\mciteSetBstMidEndSepPunct{\mcitedefaultmidpunct}
{\mcitedefaultendpunct}{\mcitedefaultseppunct}\relax
\EndOfBibitem
\bibitem[Head \emph{et~al.}(2011)Head, Briels, and Gompper]{Head2011}
D.~a. Head, W.~Briels and G.~Gompper, \emph{BMC biophysics}, 2011, \textbf{4},
  18\relax
\mciteBstWouldAddEndPuncttrue
\mciteSetBstMidEndSepPunct{\mcitedefaultmidpunct}
{\mcitedefaultendpunct}{\mcitedefaultseppunct}\relax
\EndOfBibitem
\bibitem[Pearce \emph{et~al.}(2018)Pearce, Heil, Jensen, Jones, and
  Prokop]{Pearce2018}
S.~P. Pearce, M.~Heil, O.~E. Jensen, G.~W. Jones and A.~Prokop, \emph{Bulletin
  of Mathematical Biology}, 2018, \textbf{80}, 3002--3022\relax
\mciteBstWouldAddEndPuncttrue
\mciteSetBstMidEndSepPunct{\mcitedefaultmidpunct}
{\mcitedefaultendpunct}{\mcitedefaultseppunct}\relax
\EndOfBibitem
\bibitem[Brackley \emph{et~al.}()Brackley, Morozov, and
  Marenduzzo]{BrackleyTwistable}
C.~A. Brackley, A.~N. Morozov and D.~Marenduzzo, \emph{The Journal of Chemical
  Physics}, \textbf{140}, year\relax
\mciteBstWouldAddEndPuncttrue
\mciteSetBstMidEndSepPunct{\mcitedefaultmidpunct}
{\mcitedefaultendpunct}{\mcitedefaultseppunct}\relax
\EndOfBibitem
\bibitem[de~Gennes(1979)]{DeGennes}
P.~G. de~Gennes, \emph{{Scaling Concepts in Polymer Physics }}, Cornell
  University Press, 1979\relax
\mciteBstWouldAddEndPuncttrue
\mciteSetBstMidEndSepPunct{\mcitedefaultmidpunct}
{\mcitedefaultendpunct}{\mcitedefaultseppunct}\relax
\EndOfBibitem
\bibitem[Narros \emph{et~al.}(2013)Narros, Moreno, and Likos]{Narros2013}
A.~Narros, A.~J. Moreno and C.~N. Likos, \emph{Macromolecules}, 2013,
  \textbf{46}, 3654--3668\relax
\mciteBstWouldAddEndPuncttrue
\mciteSetBstMidEndSepPunct{\mcitedefaultmidpunct}
{\mcitedefaultendpunct}{\mcitedefaultseppunct}\relax
\EndOfBibitem
\end{mcitethebibliography}
\providecommand*{\mcitethebibliography}{\thebibliography}
\csname @ifundefined\endcsname{endmcitethebibliography}
{\let\endmcitethebibliography\endthebibliography}{}

}

\newpage
\appendix

\section{Appendix A: Full Model Details}
\label{model_details}
We use the LAMMPS software~\cite{LAMMPS} to perform Brownian dynamics simulations, where the position of bead $i$ is determined by the stochastic differential equation
\begin{equation}\label{langevin}
m_i \frac{ d^2 \mathbf{r}_i }{dt^2} = -\nabla U_i + \mathbf{F}_i - \xi_i \frac{d\mathbf{r}_i }{dt} + \sqrt{2k_BT\xi_i}\boldsymbol{\eta}_i(t),
\end{equation}
where $m_i$ is the mass of the bead, $\xi_i$ is the friction it experiences due to an implicit solvent, and $\boldsymbol{\eta}_i$ is a vector representing random uncorrelated noise which obeys the following relations
\begin{equation}
\langle \eta_{\alpha}(t) \rangle = 0 ~~\mbox{and}~~  \langle \eta_{\alpha}(t)\eta_{\beta}(t') \rangle = \delta_{\alpha\beta} \delta(t-t').
\end{equation}
The potential $U_i$ is a sum of interactions between bead $i$ and all other beads, as described below. $\mathbf{F}_i$ is the active force on bead $i$ due to motor-polymer interactions. Such force, applied to the polymer by all the motors in equal measure, is directed along the polymer backbone. For simplicity we assume that all polymer and motor beads in the system have the same mass $m_i\equiv m=1$, and (except for the case of Fig.~4d) friction $\xi_i\equiv \xi=2$. The orientation of each polymer bead is described by three orthogonal unit vectors $\mathbf{\hat{f}}_i$ $\mathbf{\hat{v}}_i$ and $\mathbf{\hat{u}}_i$ which form a right-handed set of axes. A similar stochastic differential equation describes the dynamics of the orientation
\begin{equation}
I_i \frac{d \boldsymbol{\omega}_i}{dt} = \mathbf{T}_i - \xi_{r,i}\boldsymbol{\omega}_i + \sqrt{2 k_BT\xi_{r,i}}\boldsymbol{\eta}_{r,i}(t),
\end{equation}
where $\boldsymbol{\omega}_i$ is the angular velocity of bead $i$, and $\xi_{r,i}$ and $I_i$ are the rotational friction and moment of inertia respectively. The latter is set according to the Stokes-Einstein relation assuming a solid sphere. The rotational noise vector $\boldsymbol{\eta}_{r,i}(t)$ has the same properties as the translation case, and the torque $\mathbf{T}_i$ is due to the potential given below. The equations of motion are solved in LAMMPS using a standard Velocity-Verlet algorithm.
Polymer beads are connected via FENE bonds according to the potential
\begin{align}\label{FENE}
U_{\rm FENE}&(r_{i,i+1}) =& \nonumber\\ &U_{\rm WCA}(r_{i,i+1}) - 
\frac{K_{\rm FENE} R_0^2}{2} \log \left[ 1 - \left(\frac{r_{i,i+1}}{R_0}\right)^2 \right] ,
\end{align}
where $r_{i,i+1}=|\mathbf{r}_i-\mathbf{r}_{i+1}|$ is the separation of the beads, and the first term is the Weeks-Chandler-Andersen (WCA) potential
\begin{align}
\frac{U_{\rm WCA}(r_{ij})}{k_BT}  = \left\{ 
\begin{array}{ll} 
4 \left[ \left( \frac{d_{ij}}{r_{ij}}\right)^{12} - \left( \frac{d_{ij}}{r_{ij}}\right)^{6} \right] + 1, & r_{ij}<2^{1/6}d_{ij} \\
0, & \mbox{otherwise},
\end{array} \right.
\label{eq:WCA}
\end{align}
which represents a steric interaction preventing adjacent beads from overlapping. In Eq.~(\ref{eq:WCA}) $d_{ij}$ is the mean of the diameters of beads $i$ and $j$, which for simplicity is the same for motors and polymer beads, and denoted $\sigma$. In Eq.~(\ref{FENE}) $R_0$ and $K_{\rm FENE}$ are the maximum extension and strength of the bond respectively, and we use $R_0=1.6~\sigma$ and $K_{\rm FENE}=30~k_BT$ throughout. 

The bending rigidity of the polymer is introduced via a Kratky-Porod potential for every three adjacent DNA beads
\begin{equation}\label{eq:bend}
U_{\rm BEND}(\theta)=\\K_{\rm BEND} \left[ 1 - \cos(\theta) \right],
\end{equation} 
where $\theta$ is the angle between the three beads as given by
\begin{equation}
\cos(\theta) = \frac{[\mathbf{r}_i-\mathbf{r}_{i-1}]\cdot [\mathbf{r}_{i+1}-\mathbf{r}_{i}]}{|\mathbf{r}_i-\mathbf{r}_{i-1}|  |\mathbf{r}_{i+1}-\mathbf{r}_{i}|},
\end{equation}
and $K_{\rm BEND}$ is the bending energy.  In the absence of motors the persistence length in units of $\sigma$ is given by $L_p=K_{\rm BEND}/k_BT$. Unless otherwise stated we set $K_{\rm BEND}=20~k_BT$, i.e. we consider a semi-flexible polymer; this has previously been used as a model for DNA -- if we take the bead diameter $\sigma=2.5$~nm, then the persistence length (in the absence of motors) is $L_p=50$~nm, which is relevant for DNA in typical physiological conditions. Once the simulation length scale has been fixed, and using the energy unit of $k_BT$ with temperature $T=298$~K, one can map simulation time to physical units through the Brownian time $\tau_{\rm B}=\sigma^2/D$, using the Stokes-Einstein relation to find the diffusion constant for a sphere $D=k_BT/(3\pi\eta\sigma)$. For a DNA system a suitable value of fluid viscosity results in a Brownian time $\tau_B=36$ ns; with our choice of $m=1$ and $\xi=2$, the simulation time unit would then be $\tau=\tau_b/2=18$ ns.

The torque on the polymer beads is due to a potential which constrains bead orientations to lie along the polymer tangent, given by 
\begin{equation}
U_{\rm ORIENT} = K_{\rm ORIENT} \left[ 1 - \cos(\phi_i) \right],
\end{equation}
where $\phi_i$ is the angle between $\mathbf{\hat{u}}_i$ and the direction of the polymer backbone such that
\begin{equation}
\cos(\phi_i) = \frac{\mathbf{\hat{u}}_i \cdot (\mathbf{r}_{i+1}-\mathbf{r}_i) }{| \mathbf{r}_{i+1}-\mathbf{r}_i |}.
\end{equation}
We set the orientation energy $K_{\rm ORIENT}=30~k_BT$. 
Steric interactions between non-adjacent DNA beads are also given by the WCA potential [Eq.~(\ref{eq:WCA})]. Note that this is the polymer model described in Ref.~\cite{BrackleyTwistable} but without torsional rigidity.

The motors are represented by freely diffusing beads which interact with each other via a WCA potential and via the {potential given in Eq.~(\ref{motorattraction}).} {A motor is said to be ``bound'' to the polymer when its distance to a polymer bead is less than $r_{\rm m\,cut}=\sigma$.} In addition to this {interaction}, when a motor is bound to a polymer bead, it experiences an active force of magnitude $f$ which is directed along the orientation vector of the polymer bead. As a consequence of Newton's third law, the polymer bead experiences an equal and opposite force. The result is that for non-zero $f$ the motors track along the polymer substrate while driving the motion of the polymer.
{In order to avoid accumulation and jamming we want motors to detach once they have reached the end of the substrate. To achieve this we impose a large force $f_{end}\gg K_{\rm MOT}/\sigma$ to motors when bound to the last bead of the chain. Though large, this force does not appreciably influence the dynamics nor the conformation of the chain.}
Since the polymer is polar, motors can be made to move in the parallel or anti-parallel direction depending on the sign of $f$. In the present work we consider the case where all motors are the same and $f\geq0$.  


\section{Appendix B:  Motor statistics}
\label{ssec:motors}

\vspace{10pt}

\begin{table}[h!]
	\begin{center}
		\begin{tabular}{ c c c | c c c | c c c } 
			\multicolumn{3}{ c|}{$K_{\rm MOT} = 10 $}& 	\multicolumn{3}{|c|}{$K_{\rm MOT} = 20 $} & \multicolumn{3}{|c}{$K_{\rm MOT} = 30 $}\\
			$N$ & f & $\langle n \rangle$  & $N$ & f & $\langle n \rangle$ & $N$ & f & $\langle n \rangle$\\
			\hline
			100 & 6 & 9 & 200 & 6 & 165 & 200 & 6 & 200\\
			100 & 40 & 1 & 200 & 40 & 12 & 200 & 40 & 160\\
			200 & 6 & 18 & 300 & 6 & 250 & 300 & 6 & 300\\
			200 & 40 & 3 & 300  & 40 & 20 & 300 & 40 & 250\\ 
			400 & 6 & 34 & 400 & 6 & 300 & 400 & 6 & 400\\
			400 & 40 & 6 & 400  & 40 & 30 & 400 & 40 & 320 
		\end{tabular}
	\end{center} 
	\caption{Average number of motors attached $\langle n\rangle$, as measured from the simulations, for the different values of $K_{\rm MOT}$, $N$ and $f$ considered.}
	\label{table:avgnmot}
\end{table}

The average number of motors attached $\langle n \rangle$ is a key quantity, as it determines the average, total active force applied on the polymer. It is also connected to the average time a motor spends attached to the polymer, the \textit{residence time}. In equilibrium ($f = $ 0), the motors do not move along the polymer and the residence time scales as $ e^{K_{\rm MOT}/k_B T}$. 
We report the average number of attached motors $\langle n \rangle$ in Table ~\ref{table:avgnmot}, for the values of the total number of motors $N$, the interaction strength $K_{\rm MOT}$ and the motor force $f$ considered in our simulations. We observe that $\langle n \rangle$ strongly depends on $f$, as well as $N$ and $K_{\rm MOT}$. For $f\sigma/2<K_{\rm MOT}$, $\langle n \rangle$ increases with $K_{\rm MOT}$ as the residence time is still proportional to the equilibrium one; we call this the \textit{thermodynamic unbinding} regime. If $f \sigma/2 \approx K_{\rm MOT}$, then $\langle n \rangle$ decreases; we call this the \textit{activity induced unbinding} regime, and propose three possible mechanisms. First, for larger forces the motors move faster, and will ``fall off the end'' of the polymer after a time shorter than the equilibrium residence time; second, the local bends induced by the motors may provide a negative feedback, as a motor is more likely to detach as it moves around a sharp bend; and third, at very high values of $\langle n \rangle$, motor collisions might also result in a decreased residence time.


\begin{figure}[t]
	\centering
	\includegraphics{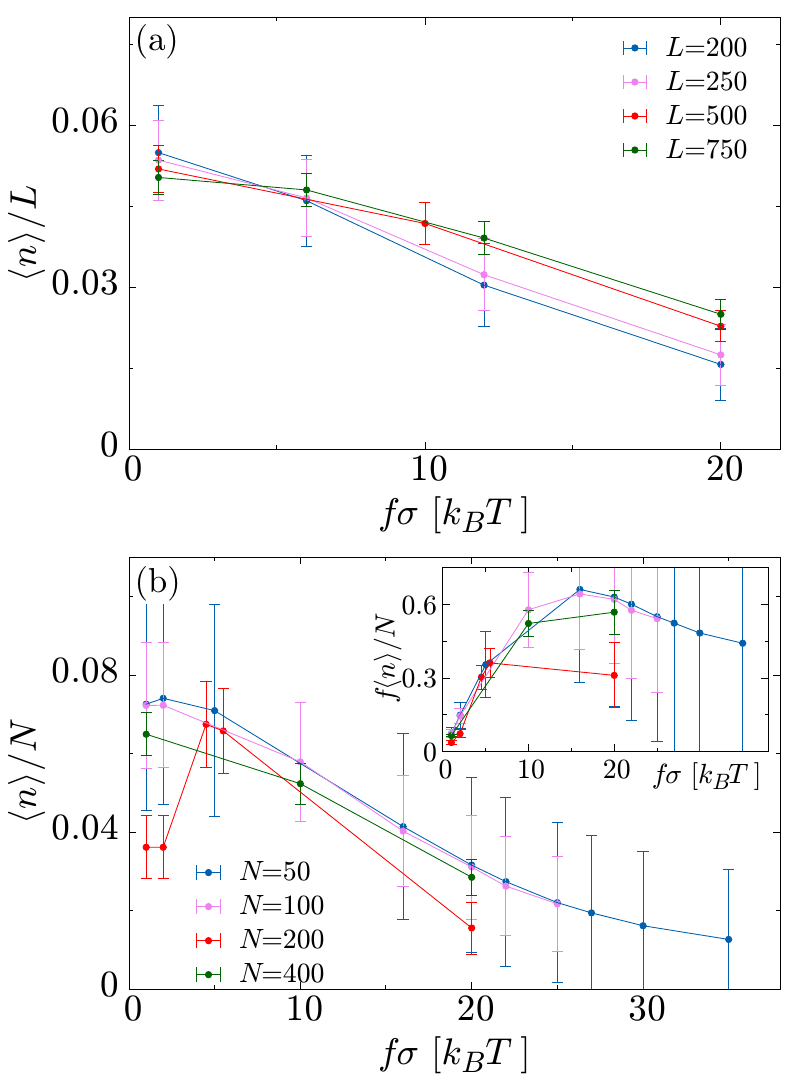}
	
	\caption{\small (a) Average fraction of occupied sites as a function of the motor force for different substrate lengths $L$ and fixed $N=$200. (b) Average fraction of bound motors for different $N$ and fixed $L=$500. Inset: Average ``used'' motor force as function of the ``nominal'' motor force for different $N$ and fixed $L=$500. In all cases, $K_{\rm MOT}=10~k_BT$.}
	\label{fig:motors}
\end{figure}


In Fig.\ref{fig:motors}a we show how the linear density of bound motors  $\langle n \rangle / L$ depends on $L$ and $f$, at constant $N =$200. For small motor force (the \textit{thermodynamic unbinding} regime) we find that $\langle n \rangle/L$ increases roughly linearly with increasing $L$, as would be expected at equilibrium; at high $f$ longer polymers are more populated than shorter ones. This suggests that motors moving off the end of the polymer play a more important role in the latter regime.  
In Fig.~\ref{fig:motors}b we plot the fraction of motors which are bound, $\langle n \rangle /N$, as a function of $f$ at fixed $L =$ 500; data for different $N$ collapse onto the same curve, implying that, even in the large $f$ regime, motor crowding does not have a large effect. Interestingly, if we plot the average ``used'' motor force $f\langle n\rangle N$ (Fig.~\ref{fig:motors}b inset) we observe non-monotonic behaviour: the total force is maximal at $f \sim$ 15 $k_B T/\sigma$.


\section{Appendix C: Scaling argument for $R_g$}\label{sec:scaling}

In the main text we have reported that the radius of gyration scales as $R_g \sim (\langle n \rangle f^2)^{-\alpha}$ with $\alpha=0.1$. We can link this observation with the scaling of $R_g$ with the contour length via the following effective blob picture: at steady state, the only length scale that can be extracted from the parameters of the system and that depend on the motor force is  $\xi_b = k_B T/f$. In analogy with traditional scaling arguments, we will take this to be the typical blob size~\cite{DeGennes}. Within a single blob the steric repulsions are not screened and the chain behaves like a self-avoiding walk (we expect the blob size to scale with the number of monomers per blob as $\sim g^{3/5}$). On a larger scale, the size of the chain will grow as $R_g \sim m^{\nu} $ where $m$ is the number of independent blobs and $\nu$ is the metric exponent. 
In order to find the number of blobs in a chain we first need to compute the number of monomers per blob, i.e.
\begin{equation}
g \sim \left( \frac{k_B T}{f} \right)^{5/3} \, .
\end{equation}
As mentioned above, the size of the polymer then scales as  
\begin{equation}
R_g \sim \xi_b\left( \frac{L}{g} \right)^{\nu} ,
\end{equation}
where $L$ is the total number of monomers in the chain. By combining these equations, we arrive at $R_g \sim f^{(5/3) \nu - 1}$. Comparing this with the numerical result from the simulations, $R_g \sim f^{-0.2}$ (Fig.~\ref{fig:bond-bond-corr}(b) inset), we obtain
\begin{equation}
\frac{5\nu}{3}-1= -0.2,
\end{equation}
or $\nu = 0.48 \approx 1/2$, i.e. the exponent of an ideal random walk. It is intriguing to notice that this value is not far from the one obtained by performing explicit simulations with different polymer lengths $L$ in Fig.~\ref{Rg}. 

\section{Appendix D: Measurements from simulated trajectories}\label{sec:shape}

As detailed in the text, the main quantities we measured from the polymer were the the radius of gyration $R_g$, the bond-bond correlation function $\beta(s)$ (from which we obtain the effective persistence length $L_p$), and the mean squared displacement of the centre of mass of the polymer. In each case the system was initiated with the polymer in a random walk conformation; the dynamics were evolved in the absence of motor interactions for $10^{6}$ time steps to reach an equilibrium polymer configuration. 
After switching on the motor interactions, the system was allowed to reach a steady state, and then quantities were determined by averaging over $3\times10^8$ time steps in each of 50 independent simulations. 

Additionally we measured two \textit{polymer shape parameters}, the asphericity and prolateness. These are calculated from the principal moments of the gyration tensor, which is defined as
\begin{equation}
S_{nm}=\frac{1}{N} \sum_{i=1}^N (r_n^{i} - R_n^{\rm cm})(r_m^i - R_m^{\rm cm} ),
\end{equation}
where $r_n^{i}$ is the $n$th element of the position vector $\mathbf{r}_i$ of the $i$th monomer, and $R_n^{\rm cm}$ is the $n$th element of the centre of mass position vector $\mathbf{R}_{\rm cm}$. Specifically the asphericity is defined
\begin{equation}
\Delta \equiv \lambda_3^2 -  \frac{(\lambda_2^2 + \lambda_1^2)}{2},
\end{equation}
and the prolateness 
\begin{equation}
S \equiv \frac{(3\lambda_1-R^2_g)(3 \lambda_2 - R_g^2)(3\lambda_3- R_g^2)}{R_g^6},
\end{equation}
where $\lambda_1>\lambda_2>\lambda_3$ are the principal moments. The radius of gyration can also be defined as $R_g^2=\lambda_1^2+\lambda_2^2+\lambda_3^2$~\cite{Narros2013}.

\end{document}